\documentclass[a4paper]{article} 
\usepackage[T1]{fontenc}
\usepackage[utf8]{inputenc}

\usepackage{authblk}

\usepackage{times}
\usepackage{fullpage}

\usepackage[nocompress]{cite}

\usepackage[british]{babel}
\usepackage{amssymb,amsmath}
\usepackage{xspace}
\usepackage[inline]{enumitem}
\usepackage{url}

\usepackage[caption=false,font=normalsize,labelfont=sf,textfont=sf]{subfig}

\usepackage[final]{listings}
\usepackage{lstcoq}
\lstnewenvironment{coqcode}[1][]
  {\lstset{basicstyle=\tt, #1}}
  {}

\usepackage[final]{listings}

\newcommand{\mytilde}{\raise.17ex\hbox{$\scriptstyle\sim$}}

\usepackage[disable]{todonotes}

\newcommand\todoXU[1]{\todo[color=blue!40, size=\tiny]{ #1}\xspace}

\newcommand\todoLR[2][]{\todo[color=gray!40, size=\tiny, #1]{{ #2}}\xspace}

\newcommand{\coq}{\textsc{Coq}\xspace}
\newcommand{\pactole}{\textsc{Pactole}\xspace}
\newcommand{\dmax}{\ensuremath{D_{\!{\textsl{max}}}}\xspace}
\newcommand{\LCM}{Look-Compute-Move\xspace}

\usepackage[pdftex]{hyperref}      
\hypersetup{%
  bookmarks         = true,
  bookmarksnumbered = true,
  bookmarksopen     = true,
  pdfpagemode       = UseNone,
  pdfborder         = {0 0 0},
  pdfstartview      = FitH,
  pdfpagelayout     = SinglePage,
  colorlinks        = true,
  urlcolor          = blue,
  linkcolor         = blue,
  pagecolor         = blue,
  anchorcolor       = blue,
  citecolor         = blue,%
  filecolor         = blue,%
  linktocpage       = true,%
  breaklinks        = true,
  backref           = true,
  pagebackref       = true,
  hyperindex        = true,
  pdftitle    = {ICDCS 2021 submission},%
  pdfsubject  = {Formal mechanized and automated framework for
    swarms of robots},%
  pdfauthor   = {},%
  pdfkeywords = {Formal Proof, Proof Assistant, Coq, Mobile Autonomous
  Robots, Distributed Algorithms, Automated Deduction, Program
  Verification}
}
\usetikzlibrary{patterns}
\pgfdeclarelayer{bg}    
\pgfsetlayers{bg,main}  

\tikzstyle{petitrond}=[circle=0pt,inner sep=1.5pt,outer sep=1.5pt,draw=black]
\tikzstyle{quadri}=[rectangle,draw]
\newcommand{\robot}[3]{\node (r) at (#1,#2) [draw,color=#3,fill,petitrond] {};}
\newcommand{\robotn}[4]{\node (#4) at (#1,#2) [draw,color=#3,fill,petitrond] {};}

\begin{document}
\title{Computer Aided Formal Design of Swarm Robotics Algorithms
\thanks{This work was partially supported by Project SAPPORO of the
  French National Research Agency (ANR) under the reference 2019-CE25-0005-1.}
}

\author[1]{Thibaut Balabonski}
\author[2]{Pierre Courtieu}
\author[1]{Robin Pelle}
\author[3]{Lionel Rieg}
\author[4]{S\'{e}bastien Tixeuil}
\author[5]{Xavier Urbain}

\affil[1]{Université Paris-Saclay, CNRS, LMF}
\affil[2]{C\'edric, Conservatoire des Artts et Métiers, Paris}
\affil[3]{VERIMAG, UMR 5160, Grenoble INP, Univ. Grenoble Alpes}
\affil[4]{Sorbonne University, CNRS, LIP6}
\affil[5]{Université Claude Bernard Lyon 1, LIRIS \textsc{umr5205}}

\date{}
\maketitle

\begin{abstract}
Previous works on formally studying mobile robotic swarms consider necessary and sufficient system hypotheses enabling to solve theoretical benchmark problems (geometric pattern formation, gathering, scattering, etc.).

We argue that formal methods can also help in the early stage of mobile robotic swarms protocol design, to obtain protocols that are correct-by-design, even for problems arising from real-world use cases, not previously studied theoretically.

Our position is supported by a concrete case study. Starting from a real-world case scenario, we jointly design the formal problem specification, a family of protocols that are able to solve the problem, and their corresponding proof of correctness, all expressed with the same formal framework. The concrete framework we use for our development is the \pactole library based on the \coq proof assistant. 
\end{abstract}

\section{Introduction}\label{sec:intro}

\subsection{Context}
Swarm robotics envisions groups of mobile robots self-organizing and cooperating toward the resolution of common objectives, such as patrolling, exploring and mapping disaster areas, constructing ad hoc mobile communication infrastructures to enable communication with rescue teams, etc. As several of those applications are life-critical, the correctness of the deployed protocols becomes of paramount importance. In turn, correctness reasoning about autonomous moving and computing entities that collaborate to achieve a global objective in a setting where unpredictable hazards may occur is complex and error prone. A first step into more formal reasoning is to use a sound \emph{mathematical model}.  
 
Suzuki \& Yamashita~\cite{suzuki99siam} introduced such a mathematical model describing the behaviour of robots in this context. The model is targeted at swarms of very weak robots evolving in harsh environments. At its core, the model simply commands individual robots to repetitively \emph{observe} their environment before \emph{computing} a path of actions to pursue and acting on it, usually by \emph{moving} to a specific location.
Three different levels of synchronization have been commonly considered. The fully-synchronous (FSYNC) case~\cite{suzuki99siam} ensures each phase of each cycle is performed simultaneously by all robots. The semi-synchronous (SSYNC) case~\cite{suzuki99siam} considers that time is discretized into rounds, and that in each round an arbitrary yet non-empty subset of the robots are active. Finally, the asynchronous (ASYNC) case~\cite{flocchini05tcs} allows arbitrary delays among the Look, Compute and Move phases, and the movement itself may take an arbitrary amount of time. 
The Look-Compute-Move model received a considerable amount of attention from the Distributed Computing community,\footnote{Yamashita received the ``Prize for Innovation in Distributed Computing'' for his seminal work on this model.} yielding a large variety of submodels induced by refined system assumptions. Such submodels were typically used to assess the solvability of a certain task assuming certain system hypotheses. As such, the Distributed Computing literature about mobile robots so far can be seen as \emph{computability-oriented}.

Alas, the various submodels make it extremely tedious to check whether a particular property of a robot protocol holds in a particular setting.  Furthermore, these variants do not behave well regarding proof reusability: checking that a property holding in a given setting also holds in another setting that is not strictly contained in the former often amounts to developing a completely new proof, regardless of the proof arguments similarity. 
This constitutes a major issue when one investigates the correctness of new solutions or implementations of existing protocols to be used in more realistic execution models. 
This problem is specially acute because of the great diversity of subtly different models: one may be tempted to simply hand-wave their way around the issue by declaring that the proof in this model is ``obviously'' also valid in this very close model, even more so as even a careful examination may not always find the most subtle errors. 
Last but not least, protocols are typically written in an informal high level language: assessing whether they conform to a particular model setting is particularly cumbersome, and may lead to hard to find mismatches.
As a result, sustained research efforts were made in the last decade to use \emph{formal methods} in the context of mobile robotic swarms.

\subsection{Related works}
Formal methods encompass a long-lasting path of research that is meant to overcome errors of human origin.
Perhaps the most well known instance in the Distributed Computing community is the~\emph{Temporal Logic of Actions} and its companion tools TLA/TLA+\cite{lamport1994ACM,cousineau12fm}. Though very expressive, TLA is designed for the shared
memory and message passing contexts, thus not perfectly suited to studying mobile robotic swarms.
\emph{Model-checking} and its powerful automation proved useful to find bugs in existing literature~\cite{berard16dc,doan16sofl,doan17opodis}, and to assess formally published algorithms~\cite{devismes12sss,berard16dc}, in a simpler setting where robots evolve in a \emph{discrete space} where the number of possible positions is finite. Automatic program synthesis (for the problem of perpetual exclusive exploration in a ring-shaped discrete space) is due to Bonnet \emph{et al.}~\cite{bonnet14wssr}, and can be used to obtain automatically algorithms that are ``correct-by-design''. The approach was refined by Millet \emph{et al.}~\cite{MPST14c} for the problem of gathering in a discrete ring network. 
However, those approaches are limited to instances with few robots. Generalizing them to an arbitrary number of robots with similar models is doubtful as Sangnier \emph{et al.}~\cite{sangnier20fmsd} proved that safety and reachability problems are undecidable in the parameterized case. Another limitation of the above approaches is that they \emph{only} consider cases where mobile robots \emph{evolve in a discrete space} (\emph{i.e.}, graph). This limitation is due to the model used, that closely matches the original execution model by Suzuki and Yamashita~\cite{suzuki99siam}. As a computer can only model a finite set of locations, a continuous 2D Euclidean space cannot be expressed in this model. Défago et al.~\cite{defago20srds} used a more abstract model to model-check rendez-vous algorithms in a continuous 2D Euclidean space, however, their model is highly specific to rendez-vous and thus is not as versatile as one could hope, hinting at a more general and systematic technique.
 
The approach on which we focus in this work is \emph{formal proof}, that is proof development mechanically certified by a proof assistant. 
Mechanical proof assistants are proof management systems where a user can
express data, programs, theorems and proofs. In sharp contrast with automated
provers (like model-checkers), they are mostly interactive, and thus require some kind of expertise from their users. Sceptical proof assistants provide an additional guarantee by checking mechanically the soundness of a proof after it has been interactively developed.
Formal proof allows for more genericity as this approach is not limited to \emph{particular instances} of algorithms. 
During the last twenty years, the use of tool-assisted verification has extended
to the validation of distributed processes, in contexts such as process algebras~\cite{fokkink07,bezem97fac}, symmetric interconnection networks~\cite{gaspar14ijpp}, message passing settings~\cite{kuefner12ifiptcs}, and self-stabilization~\cite{deng09tase,altisen16forte}, etc.
The main approach for mechanized proof dedicated to swarms of mobile
entities is so far the \pactole\footnote{\tt \url{https://pactole.liris.cnrs.fr}} framework.
Initiated in 2010, The \pactole framework enabled the use of
high-order logic to certify impossibility results, as well as
soundness of protocol, for swarms of autonomous mobile robots.  To
certify results and to guarantee the soundness of theorems, the proof
assistant it uses is \coq.
Briefly, \coq is a Curry-Howard-based interactive proof assistant that
enjoys a trustworthy kernel. Its base language is a very expressive
$\lambda$-calculus, the \emph{Calculus of Inductive Constructions}
\cite{coquand90colog}, where datatypes, objects, algorithms,
theorems and proofs can be expressed in a unified way, as terms. The
syntax is close to that of an ML-like programming language, and a proof
development consists in trying to build, interactively and using
tactics, a $\lambda$-term, the type of which corresponds to the theorem
to be proven (Curry-Howard style).  The small kernel of \coq simply
type-checks $\lambda$-terms to ensure soundness.  Most importantly :
\emph{a theorem or a lemma can only be saved/defined in the system if
  it comes with its type-checked proof}.
Designed for mobile entities, and making the most of \coq's assets,
\pactole allows for working on a given protocol to establish and
certify its correctness \cite{courtieu16disc,balabonski19tocs}, as
well as for quantifying over all protocol so as to prove
\emph{impossibility} results
\cite{auger13sss,courtieu15ipl,balabonski18icdcn}, with an unspecified
number of robots, possibly including a proportion of Byzantine faults, in
continuous or discrete spaces.
FSYNC/SSYNC and ASYNC modes are all supported, and the framework is
expressive enough to state and certify formally results as theoretical
as comparisons between demons or models \cite{balabonski2019netys}.

\subsection{Our Contribution}
\label{sec:contribution}

Taking some perspective over aforementioned works mixing formal methods and swarm robotics, one can only notice that the computability-centric approach of the Suzuki and Yamashita model yielded a concentration of efforts towards few benchmark\todoXU{on \textbf{a} few ?} problems that are theoretically interesting (one can get impossibility results or correctness certification) but of little practical relevance, such as perpetual or terminating exploration of a ring-shaped graph, and gathering or concentrating all robots at a particular location.

On the other side, relevant practical problems, such as constructing ad hoc mobile communication infrastructures to enable communication with rescue teams, remain untouched using a formal approach. Yet, their correctness is crucial, and possibly life-critical, so it should be assessed formally and mechanically verified. Overall, for those practical problems, the question is not really to characterize which system hypotheses enable problem solvability, but rather how to design a provably correct solution using hypotheses that correspond to real devices.

This paper is the first step in this direction. In more details, we start from a real-life application scenario to jointly design \emph{(i)} its formal specification, \emph{(ii)} a family of protocols that are able to solve the problem, and \emph{(iii)} their corresponding proof of correctness, all expressed with the same formal framework, \pactole. In this process, we illustrate how formal methods and \pactole in particular could be used to derive protocols that are correct-by-design before they are deployed to actual devices.  

The developments for \coq v8.12 described in this work are available
at {\small \tt \url{https://pactole.liris.cnrs.fr/pub/connection_8.12.tgz}}
\section{Search and Rescue, the life line maintainance problem}\label{sec:problematique}

One of the most advertised applications of robotic swarms is the Search
and Rescue situation: a devastated zone has to be explored to bring
assistance to survivors. A particular scenario in that context is the
``life line'' one where, static communication infrastructures having
been destroyed, a robotic swarm establishes a dynamic network between
a moving search team and the rescue station, by sending mobile
transmitters relays.

The same problematic is faced in the less dramatic context
of the pursuit of a hornet with drones whose sensors and transmitters
are of limited range \cite{reynaud16ahn}.

In such situations :
\begin{itemize}
\item A mobile target follows an unpredictable path and must always be
  in sight.
\item Mobile robots with limited sensors/transmitters track and follow
  the target.
\item To maintain contact with the base station which might be lost
  due to range limits, some of the robots act as relays.
\end{itemize}
The link between the base and the target, the so-called life line,
must never be broken. More precisely: the existence of such a life
line should be ensured at each and every point in the execution.

As practical applications may be critical, with lives at stake, it is
imperative for that invariant to hold, and one needs formal guarantees
about it.

\subsection{Space}

Focusing on the search and rescue scenario, we
have drones flying and monitoring a rescue team on the ground. Though
the most natural space would be the 3D space, it is enough to consider
that the mobile robots are moving on a continuous plane, at a fixed altitude.
Thus, we can choose the space to be the 2D Euclidean plane.\footnote{This incidentally will allow us to use vertical movement as a way to remove robots from the protocol, see Section~\ref{sec:removal}.}

In this space, there is a 
point called the \emph{base} from which robots are launched.  %
The second defining object is the \emph{companion} 
which moves on the plane and follows the rescue team, regardless of
other robots; it has to be
connected to the base, one way or another. %

\subsection{Basic robot characteristics and sensors}
To avoid the problem being trivially solvable, we require the sensors
and transmitters to have a limited range.  %
Hence we grant robots the ability to see neighbouring robots up to
some \emph{finite range} \dmax.  %
Indeed, infinite range would make the whole protocol moot as a single
robot located at the base would be able to see the target. %

Note that we merge vision and transmission capabilities. As a matter of
fact relays that do not see others would probably be lost, and the
point of a relay is to have neighbours to transmit to. %

There is no need to consider multiplicities, as the protocol has to
avoid collision: two launched robots will never be at the same
location (provided this property holds in the starting
configuration).

Similarly, to avoid any trivial counter-example involving a straight
line fleeing of the companion following the search team, we constrain
the speed of drones to a certain $D$ denoting the bound on the
distance that can be travelled within one cycle. 
Also, we only consider executions where the number of available relays
is sufficient. 

\subsection{Mode of synchronicity, movements, and  initial configuration}

We may assume the execution to be fully synchronous, with a short time
span for each cycle.
As pointed out in Section~\ref{sec:contribution}, our goal is \emph{not} to find minimal synchrony assumptions that allow problem solvability, but rather make sensible and realistic assumptions that permit deployment with actual devices. When robots are homogeneous, they operate at a similar pace, and the fully synchronous model is a sensible choice. Furthermore, keeping the time span short makes the assumption close to what continuous time practitioners envision~\cite{kling19bookchapter,castenow20spaa}.

For the same reason, we also consider movements to be rigid, that is robots do not start a new cycle before they have reached their computed destination. Since robots always select as a target another robot within the visibility radius, movements are always of limited length.

The fact that the rescue team departs from the base allows us to
define a notion of \emph{valid} initial configuration. Trying to solve
the problem if the rescue team is already out of reach would make no
sense in that context. 

\subsection{Informal specification of the problem}
As it is critical to keep the rescue team connected to the base, we
want a formal guarantee that there is always a chain of robots going from the base to the
target, such that robots along this chain can relay
communication. 

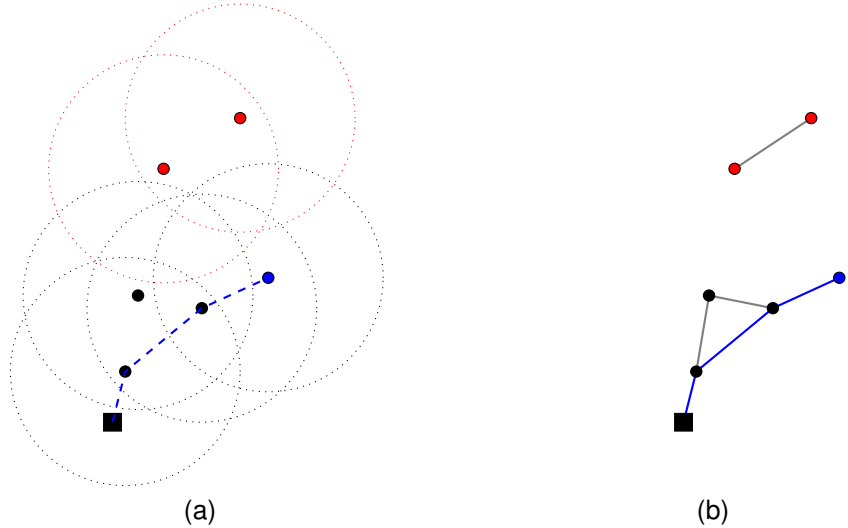
\begin{figure}
\centering
\subfloat[][]{\begin{tikzpicture}[scale=0.56]
        \begin{scope}[scale=0.3]
          \robot{8.2}{7.4}{blue}
          \node[quadri][fill=black] (base) at (-4,-4){}; 
          \draw[dotted] (8.2,7.4) circle(9);
          
          \robot{3.0}{5.0}{black}
          \draw[dotted] (3,5) circle(9);
          \robot{-3}{0}{black}
          \draw[dotted] (-3,0) circle(9);
          \draw[dashed, color=blue, thick] (-4,-4) -- (-3,0) -- (3,5) -- (8.2,7.4);   
          \robot{0}{16}{red}
          \robot{6}{20}{red}
          \robot{-2}{6}{black}
          \draw[dotted] (-2,6) circle (9) ;
          \draw[dotted, color = red] (0,16) circle(9);
          \draw[dotted, color = red] (6,20) circle(9);
        \end{scope}
      \end{tikzpicture}
} %
\hfil %
\subfloat[][]{
  \begin{tikzpicture}[scale=0.56]
       \begin{scope}[scale=0.3]
          \draw[dotted, color=white] (-3,0) circle(9);
          \draw[color=blue, thick] (-4,-4) -- (-3,0) -- (3,5) -- (8.2,7.4);   
          \draw[thick, color=gray] (0,16)--(6,20);
          \draw[thick, color=gray] (-3,0) -- (-2,6) -- (3,5);
          \robot{8.2}{7.4}{blue}
          \node[quadri][fill=black] (base) at (-4,-4){}; 
          \robot{3.0}{5.0}{black}
          \robot{-3}{0}{black}
          \robot{0}{16}{red}
          \robot{6}{20}{red}
          \robot{-2}{6}{black}
        \end{scope}
      \end{tikzpicture}
}   \caption{Let us consider in (a) a configuration of the system where
    \protect\begin{tikzpicture}\protect\node[quadri][fill=black] (base)
      at (-4,-4){}; \protect\end{tikzpicture} is the base,
    \protect\begin{tikzpicture}\protect\robot{8.2}{7.4}{blue}\protect\end{tikzpicture}
    is the companion, and where visibility range are denoted as
    dotted circles. In this situation, the two robots
    \protect\begin{tikzpicture}\protect\robot{8.2}{7.4}{red}\protect\end{tikzpicture}
     cannot detect the others. Nonetheless there is a chain of
    visibility between the base and the companion, denoted as a
    dashed blue path.   
    The corresponding visibility graph is in (b), with a suitable path we ask
    for in blue.
  }
  \label{fig:visgraph}
 \end{figure}

Given a configuration of the system (see \emph{e.g.} Figure~\ref{fig:visgraph}), we say that two robots are
\emph{connected} if they are located at most \dmax apart, that is
within visibility range. 
We formalize this idea by a \emph{visibility graph}, where nodes are robots and edges relate connected robots.
Thus, having a chain of connected robots going from the base to the companion can be expressed as \emph{having a path from the base to the companion in the visibility graph}.

A second objective of our protocol is to avoid collision for obvious
reasons, that is, to make sure that no two distinct robots are ever at the same location.

Overall, we end up with:
\begin{itemize}
\item A list of hypotheses characterizing our context and the
  environment, on space, sensors, synchronicity\dots
\item An invariant that, provided there are enough relays to
  be sent, 
  has to hold at each point of any
execution starting from a valid configuration: there is a path in the
graph of visibility and there is no collision. 
\end{itemize}

\section{A Brief Overview of the \pactole Framework}
\label{sec:pactole}

One of the aims of \pactole is to stay as simple and as close as possible to the definitions of the robotic swarm community.

Thus, a state of the overall system, a \emph{configuration} is defined as the collection of the states of all robots, conveniently combined into a map from robot names to robot states:
\begin{coqcode}[gobble=2]
  Definition configuration := ident -> state.
\end{coqcode}
A robot \emph{state} can be anything (to accurately describe reality) but must at least contains its \emph{location}, accessible through a function \lstinline{get_location : state -> location}, where the type \lstinline{location} denotes the space where robots evolve.

An \emph{execution} is an infinite sequence of configurations:
\begin{coqcode}[gobble=2]
  Definition execution := Stream.t configuration.
\end{coqcode}
Executions are usually built by executing a protocol (called a \emph{robogram}) against an environment, represented as a \emph{demon}, that is, an infinite sequence of decisions called \emph{demonic actions}.
\begin{coqcode}[gobble=2]
  Definition demon := Stream.t demonic_action.
\end{coqcode}
The robogram represents the Compute part of the \LCM cycle.
It takes the observation of the robot as input and outputs the action that the robot should perform.
\begin{coqcode}[gobble=2]
  Definition robogram := observation -> action.
\end{coqcode}

This \emph{observation} denotes a degraded version of the configuration centred on the observing robot, depending on its sensors.
It is a parameter of the model and its computation from a (local) configuration is performed by an \lstinline{obs_from_config} function, which hides the information unavailable to robots and takes as input the configuration and the state of the observing robot.
This function is specified by a logical formula \lstinline{obs_is_ok} relating any configuration to its possible observation from any robot state.
\begin{coqcode}[gobble=2]
  Parameter observation : Type.
  Parameter obs_from_config :
    configuration -> state -> observation.
  Parameter obs_is_ok :
    observation ->  configuration -> state -> Prop.
  Parameter obs_from_config_spec : forall config st,
    obs_is_ok (obs_from_config config st) config st.
\end{coqcode}

To represent the fact that robots observe from a personal point of view, they have their own \emph{frame of reference} that need not be consistent in time or with other robots (other orientation, other scale, other origin, etc.).
This frame of reference allows to create a \emph{local configuration}
(by opposition to the point of view of the demon denoted as \emph{global}) from which the observation is computed, it depends on the underlying space and it is picked by the demon.

In such an execution, the robogram corresponds to one \LCM cycle
and the demonic action to the reaction of the environment.
Their interaction is described by a function \lstinline{round} so that
the resulting execution is simply repeatedly calling this function
with the robogram, the demon and the starting configuration. 

The \lstinline{round} function is the heart of the model, implementing the Look-Compute-Move cycle and computing the configuration obtained after one round.
Note that this function is the same for all variants, FSYNC/SSYNC/ASYNC synchronization, all spaces, all sensors, etc.
This is done in the following consecutive steps for each robot name \lstinline{id}:
\begin{enumerate}
  \item If the robot \lstinline{id} is not activated, its state may undergo some change by the \lstinline{inactive} function to represent an ongoing action or the effect of the environment.
  \item If \lstinline{id} is a byzantine robot, it is relocated by the demonic action \lstinline{da}.
  \item Use the local frame of reference provided by \lstinline{da} to compute the local configuration.
  \item Transform this local configuration into an observation.
  \item Apply the robogram on this observation.
  \item If moves are flexible, compute the new position of \lstinline{id} using information given by \lstinline{da}.
  \item Convert the new position from the local frame to the global one.
\end{enumerate}

The full \lstinline{round} function is:
\begin{coqcode}
Definition round r da conf : configuration :=
  (* for a given robot, we compute the new configuration *)
 fun id =>
  let state := conf id in (* id's state read from conf *)
   (* first see whether the robot is activated *)
  if da.(activate) id
  then match id with
   | Byz b => (* byzantine robots *)
       da.(relocate_byz) conf b
   | Good g =>
     (* change the frame of reference *)
     let frame_choice := da.(change_frame) conf g in
     let new_frame :=
        frame_choice_bijection frame_choice in
     let local_config := map_config new_frame conf in
     let local_state := local_config (Good g) in
     (* compute the observation *)
     let obs :=
        obs_from_config local_config local_state in
     (* apply r on the observation *)
     let local_robot_decision := r obs in
     (* the demon chooses how to update the state *)
     let choice :=
        da.(choose_update) local_config g
                           local_robot_decision in
     (* the actual update by the update function *)
     let new_local_state := update local_config g
       frame_choice local_robot_decision choice in
     (* return to the global frame of reference *)
     new_frame $^{-1}$ new_local_state
     end
  else inactive conf id (da.(choose_inactive) conf id).
\end{coqcode}

\section{Co-Designing Specifications, Solutions, and Proofs within the Formal Framework}\label{sec:pasapas}

The first step is to instantiate in the formal framework the
assumptions listed in Section~\ref{sec:problematique}, as well as the property
that we want to be holding throughout a relevant execution.

Those specifications are ``refined'' until a set of constraints on
robots and protocol candidates is formally proven sufficient.
Ultimately, our family of protocols will be given by an abstract protocol parameterized by several functions that will be specified enough for the correctness proof to hold.

The final specifications are thus designed together with a solution and its proof.

For the sake of readability we use \lstinline!cf x! as a shorthand for
\lstinline!cf (Good x)!, i.e. the position of a (non-Byzantine) robot
named \lstinline!x! in a configuration
\lstinline!cf!.\footnote{For the sake of clarity, some notations may slightly
  differ between the actual code and this section, and some
  irrelevant technical overhead may have been pruned.}

\subsection{Specifications}\label{sec:spec}
The parameters of the problem are the number $n$ of robots, the
visibility radius \dmax of robots and the maximum distance $D$ they
can travel in a round.
\begin{coqcode}[gobble=2]
  Parameter n : nat.
  Parameters D Dmax : R.
\end{coqcode}

Using the number $n$ of robots, we define canonical names to be able to refer to them:
\begin{coqcode}[gobble=2]
  Instance Robot_Names : Names := Robots n 0.
\end{coqcode}
The \lstinline{0} represents the absence of Byzantine failures.

The space is the 2D Euclidean space, already predefined in \pactole:
\begin{coqcode}[gobble=2]
  Instance Loc : Location := make_Location R2.
\end{coqcode}

Initially, a robot's state contains only its location, and its observation consists of the set of inhabited locations within its vision range:
\begin{coqcode}[gobble=2]
  Instance first_State : State R2 := OnlyLocation.
  Instance SetObs : Observation :=
    LimitedSetObervation.limited_set_observation Dmax.
\end{coqcode}

The changes of frame of reference allowing to switch between
global 
and local observation (and vice-versa) are similarities (that is, moving the origin around and changing the orientation, chirality, and scale):
\begin{coqcode}[gobble=2]
  Instance first_Frame : frame_choice (similarity location)
    := FrameChoiceSimilarity.
\end{coqcode}

The FSYNC setting is reflected in an hypothesis made over demonic
actions, by requiring property \lstinline{FSYNC_da} to hold (as will be seen the final
statement of the invariant).

Movements are defined as rigid.
\begin{coqcode}[gobble=2]
  Instance setting_is_rigid : RigidSetting.
\end{coqcode} 

As a consequence of our FSYNC and rigid setting with no Byzantine
failure, 
some parameters of the \pactole framework are not used and can be set to arbitrary values, namely what happens to a robot whenever it is inactive, how the demon interferes with a robot movement, what happens to robot suffering Byzantine failures, etc.  
Essentially, one can skip the first two steps of the \lstinline{round} function described in Section~\ref{sec:pactole} and focus on the \LCM cycle.

\begin{coqcode}[gobble=2]
  Context {inactive_choice_ila : inactive_choice bool}.
  Instance demon_upd : update_choice unit := NoChoice.
\end{coqcode} 

We say of a robot participating to the task that it is
\emph{launched} (by the base station) and \emph{alive}.
(These notions will be formally motivated and introduced during the design of the protocol, respectively in Sections~\ref{sec:base} and~\ref{sec:removal}.) \todoXU{J'ajoute ça pour limiter les dégâts} \todoLR{pareil}
The main goal of any candidate protocol is to maintain the following properties
at any moment during the execution:

\begin{enumerate}
\item  There is no collision between robots, i.e. any two
  distinct alive robots not at base do not share the same location:
  \begin{coqcode}[gobble=4]
    Definition no_collision_conf (cf : config) :=
      forall g g', g <> g'
        -> get_launched (cf g) = true
        -> get_launched (cf g') = true
        -> get_alive (cf g) = true
        -> get_alive (cf g') = true
        -> dist (get_loc (cf g)) (get_loc (cf g')) <> 0%R.
  \end{coqcode}

\item There is a sequence of connected robots from the base
  to the companion (numbered $0$, always alive). In other words: for any
  robot alive (we shall see that there is always some robot alive on base) either it is
  the companion or it has a visible, active, launched neighbour with
  a smaller id.

\begin{coqcode}
Definition path_conf (cf:config) := forall g,
 get_alive (cf g) = true
 -> get_ident (cf g) = 0
   \/ exists g',
       dist (get_loc (cf g)) (get_loc (cf g')) <= Dmax
       /\ get_alive (cf g') = true
       /\ get_launched (cf g') = true
       /\ get_ident (cf g') < get_ident (cf g).
\end{coqcode}
\end{enumerate}

Note that this definition implies the existence of a connection from the
base to the companion only if there is a robot alive (launched or not) close
to the base, i.e. the set of robots waiting to be launched never gets
exhausted. The fact that this property always holds during the considered
execution is stated as the premise \lstinline{exists_at_based} below.

The preservation of this invariant is expressed as
\begin{coqcode}
Definition NoCollAndPath e :=
  Stream.forever (Stream.instant
(*@\hfill@*)(fun cf => no_collision_conf cf/\path_conf cf)) e.
\end{coqcode}

Being assumed that the invariant holds in the initial configuration \lstinline{config_init} 
The final lemma will state that the invariant holds forever along any execution of the
(abstract and parametric) 
protocol \lstinline!rbg_ila!, provided that 
  the set of robots alive at base never
gets empty. 
\begin{coqcode}
(* There is a robot not yet launched *)
Definition exists_at_base cf := exists g, get_launched (cf g) = false.
(* This property holds forever *)
Definition exists_at_based e :=
  Stream.forever (Stream.instant (exists_at_base)) e.
(* Hypotheses on each demonic action: it is FSYNC
   and local frames are centred on the observing robot *)
Definition da_assumption da := change_frame_origin da /\ FSYNC_da da.
(* The demon must forever satisfy these properties *)
Definition demon_ILA demon :=
  Stream.forever (Stream.instant da_assumption) demon.
(* Main lemma *)
Lemma validity_conf_init:
  forall demon, demon_ILA demon
    -> exists_at_based (execute rbg_ila demon config_init)
    -> NoCollAndPath (execute rbg_ila demon config_init).
  \end{coqcode}

  We show in the remainder of this work that the aforementioned lemma
  can in fact be formally proven for any \lstinline!rbg_ila!
  fulfilling some sufficient conditions that we discovered in the next
  section. A concrete suitable protocol is outlined 
  in Section \ref{sec:extr-sample-solut}.

\subsection{The March towards a Family of Solutions}\label{sec:raffinement}

We describe in the following how we refine this initial formal model according
to the issues we ran into and the hypotheses (whether necessary or by design)
we find helpful along the protocol and proof co-development.
We update the formal instantiation 
accordingly. For the sake of clarity, we use the following notation:
the successive re-definitions of the parameters are indexed with the
refinement step (e.g.  \lstinline!State$_2$! is second refinement 
of the initial default robot state
\lstinline!State$_0$!). We drop that numbering when we reach the final
version of the parameters.

\newcounter{mvers} 
\setcounter{mvers}{0}
\newcommand{\svers}{\arabic{mvers}}

\subsubsection{Robot path needs an orientation}\label{sec:id}~
\stepcounter{mvers}

\noindent
\textbf{Problem}: When robots see neither the base nor the companion,
for instance when they only see their two neighbours on the life
line, 
they have no way of knowing in which direction the base lies and in
which direction the companion lies.

This information is important since the behaviour of relay robots is not symmetric: they should follow the companion as the base is responsible for launching new relay robots whenever necessary.
Therefore relay robots need a way to know in which direction lies the companion.

\noindent
\textbf{Solution}: We provide visible identification and a strict
ordering between those robot identifiers. To this goal we use a
unique positive integer per robot, and assume that robots will be
launched in ascending order.
This way, the direction of the companion is given by smaller identifiers.
The definition of a state is updated accordingly.

Note that the newly introduced \emph{identifiers} are visible
information and are not to be
confused with \lstinline{Names} of robots (Section~\ref{sec:spec}), which are constructs
internal to the formal framework. 

\begin{coqcode}
Definition identifier := nat.
Definition info$_\svers$ := identifier.
Definition State$_\svers$ := AddInfo info$_\svers$ OnlyLocation.
\end{coqcode}

The companion will be given number $0$, and at any moment if a robot
is launched 
then all other robots already launched have a smaller id.

\subsubsection{Distinguishing robots waiting at base}\label{sec:base}~
\stepcounter{mvers}

\noindent
\textbf{Problem}: What if a robot is located at the base but is already launched? Should
this configuration be indistinguishable from the one where it is not
yet launched? Obviously not, as collisions may occur in the former but not in the latter.

\noindent
\textbf{Solution}: The state of a robot should contain a
boolean indicating if the robot is ``launched'' or not.
The state should be updated accordingly.

\begin{coqcode}
Definition launched := bool.
Definition info$_\svers$ := identifier*launched.
Definition State$_\svers$ := AddInfo info$_\svers$ OnlyLocation.
\end{coqcode}

\subsubsection{Robots may have to withdraw from the system}\label{sec:removal}~
\stepcounter{mvers}

\noindent
\textbf{Problem}: When the companion comes closer to the base, the
connection line shrinks making the connecting robots
possibly unable to stay out of collision risk. 
Therefore a useful feature of a candidate protocol would be the
ability to withdraw from the line.

\noindent
\textbf{Solution}: We assume available any arbitrary procedure for robot
removal, and we add a boolean \lstinline!alive! to the robot state,
making sure that dead (that is, withdrawn) robots are not taken into consideration by alive robots.\footnote{For instance, we may make them change altitude making sure they no longer represent a collision risk.}
For simplicity and separation of concern, we make dead robots not even observable by the other robots by adding new constraints to the specification of the observation.

\begin{coqcode}
Definition alive := bool.
Definition info$_\svers$ := identifier*alive*launched.
Definition State$_\svers$ := AddInfo info$_\svers$ OnlyLocation.
Definition obs_is_ok s config pt:
 get_alive pt = true
 /\ (forall l, In l s <-> ... /\ get_alive l == true).
\end{coqcode}

\subsubsection{Pruning the space of candidate solutions}\label{sec:lowerid}~

\noindent
Since robots are ordered, and the problem essentially revolves around following some robot of lower identifier, we choose to consider protocols where any given robot makes its decisions depending only on the robots of lower identifier. While this choice is not an answer to any particular problem, it reduces drastically the design space we explore. Once we make this assumption, several formerly open choices become closed. In particular, no robot of small identifier will ever take into account robots of higher identifier nor try to avoid collision with them. That means avoiding collisions and, if necessary, withdrawing, will always be the responsibility of the robots with higher identifiers.

\subsubsection{Robots need to warn neighbours before withdrawing}\label{sec:light}~
\stepcounter{mvers}

\noindent
\textbf{Problem}: A robot too close to others shall disappear, but
there are cases where a robot immediately behind it may move too far
and break the link, 
precisely \emph{because it tried to avoid that collision}, even if that
collision is now impossible due to the robot removal. 
This situation is illustrated on figure~\ref{fig:lighta}.
We decide to give robots the ability to warn neighbours about the
possibility that it may withdraw in the next round.

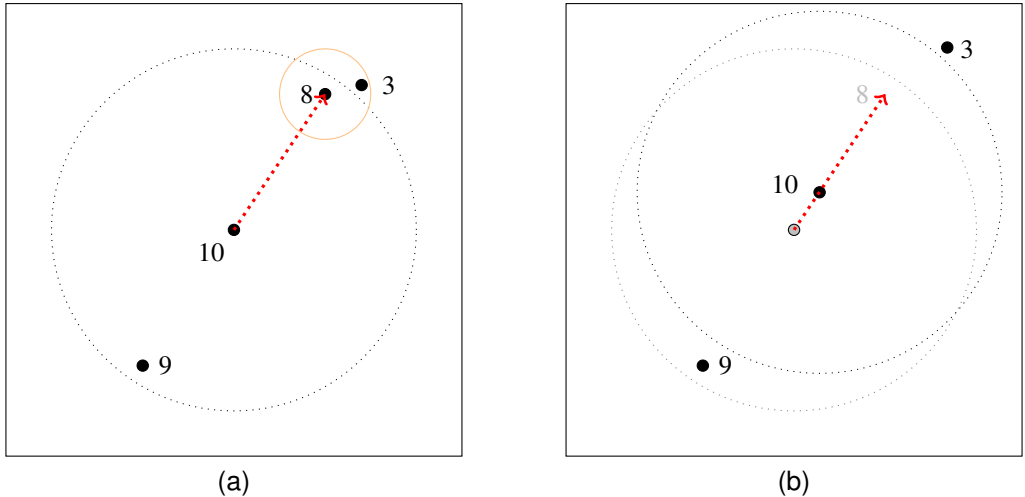
\begin{figure}
\centering
  \subfloat[][]{
  \begin{tikzpicture}
    \begin{scope}[scale=.6]
      \draw (-5,-5) rectangle (5,5);
      \robot{0}{0}{black}
      \node (r1) at (-.5,-.5) {10};
      \node (r2) at (1.6,3) {8};
      \node (r3) at (3.4,3.2) {3};
      \node (r4) at (-1.5,-3) {9};
      \robot{-2}{-3}{black}
  \draw[dotted] (0,0) circle(4);
  \robot{2}{3}{black}
  \draw[color=orange!50] (2,3) circle(1);
  \robot{2.8}{3.2}{black}
  \draw[->, dotted, very thick, color=red] (0,0) -- (2,3);
  \end{scope}
\end{tikzpicture}
}%
\hfil
\subfloat[][]{
\begin{tikzpicture}
    \begin{scope}[scale=.6]
            \draw (-5,-5) rectangle (5,5);
      \robot{0}{0}{gray!50}
  \robot{0.56}{0.83}{black}
  \draw[dotted] (0.56,0.83) circle(4);
  \draw[dotted, color=gray] (0,0) circle(4);
  \node (dead) at (2,3) {};
      \node (r1) at (-.2,1) {10};
      \node[color=gray!50] (r2) at (1.5,3) {8};
      \node (r3) at (3.8,4) {3};
  \draw[->, dotted, very thick, color=red] (0,0) -- (2,3);
  \robot{2.8+.56}{4.03}{black}
      \robot{-2}{-3}{black}
  \node (r4) at (-1.5,-3) {9};
  \end{scope}
\end{tikzpicture}
}%
\caption{In situation (a), robot 8 will withdraw as it is too close to
  3. Robot 10 can see 8 and 9, 3 is out of sight, and it may choose 8
  as its target. Situation (b) shows what can happen in this case: 10
  moves towards 8 (its previous position is shown in grey), which
  disappears. With 3 moving further, 10's connection with any suitable robot is
  lost.}\label{fig:lighta}
\end{figure}
\begin{figure}
  \ContinuedFloat
  \addtocounter{figure}{1} 
  \centering
\subfloat[][]{
  \begin{tikzpicture}
    \begin{scope}[scale=.6]
      \draw (-5,-5) rectangle (5,5);
      \robot{0}{0}{black}
      \node (r1) at (-.5,-.5) {10};
      \node (r2) at (1.6,3) {8};
      \node (r3) at (3.4,3.2) {3};
      \node (r4) at (-1.5,-3) {9};
      \robot{-2}{-3}{black}
  \draw[dotted] (0,0) circle(4);
  \robot{2}{3}{red}
  \draw[color=orange!50] (2,3) circle(1);
  \robot{2.8}{3.2}{black}
  \draw[->, dotted, very thick, color=red] (0,0) -- (-2,-3);
  \end{scope}
\end{tikzpicture}
}
 \hfil
\subfloat[][]{
  \begin{tikzpicture}
    \begin{scope}[scale=.6]
          \draw (-5,-5) rectangle (5,5);
      \robot{0}{0}{gray!50}
  \robot{-0.56}{-0.83}{black}
  \draw[dotted] (-0.56,-0.83) circle(4);
  \draw[dotted, color=gray] (0,0) circle(4);
  \node (dead) at (2,3) {};
      \node (r1) at (.2,-1) {10};
      \node[color=gray!50] (r2) at (1.5,3) {8};
      \node (r3) at (3.8,4) {3};
  \draw[->, dotted, very thick, color=red] (0,0) -- (-2,-3);
  \robot{2.8+.56}{4.03}{black}
      \robot{-2}{-3}{black}
  \node (r4) at (-1.5,-3) {9};
  \end{scope}
\end{tikzpicture}
}
\caption{Situation (c) is similar to Figure~\ref{fig:lighta} (a) as 8 will withdraw but now it \emph{switches its light
    on} (here in red). As 10 knows that 8 is likely to disappear, it
  chooses 9 as target, thus keeping a connection with a robot of lower
  identifier (situation (d)).}\label{fig:lightb}
\end{figure}
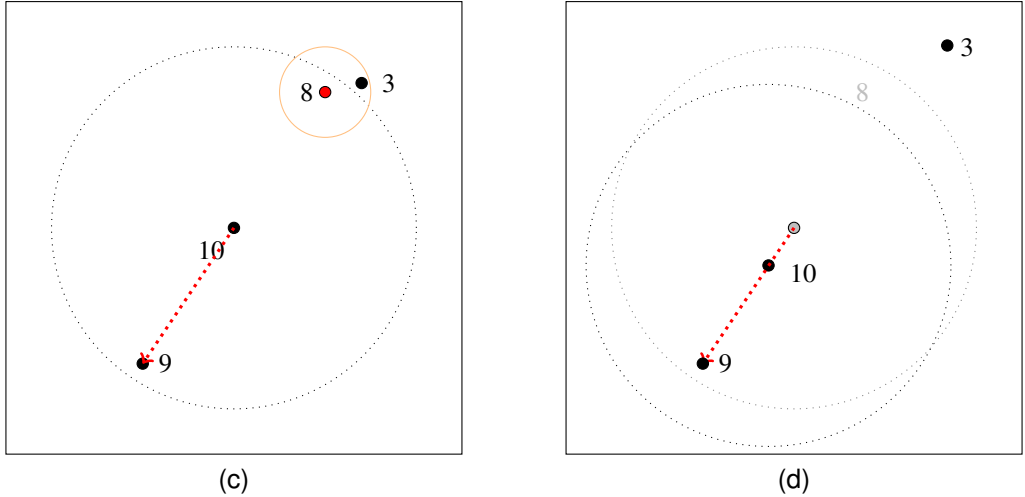

\noindent
\textbf{Solution}: We add a \emph{light} to the robots capabilities.
The light of a robot is either \lstinline{on} or \lstinline{off}.
Any robot that sees another robot sees its light.
Light \lstinline{on} means the robot might be about to withdraw and should not
be followed anymore, see Figure~\ref{fig:lightb}.

\begin{coqcode}
Definition light := bool
Definition info$_\svers$ := identifiant*light*alive*launched.
Definition State$_\svers$ := AddInfo info$_\svers$ OnlyLocation.
\end{coqcode}

\subsubsection{A robot cannot predict the moves of other robots}

\noindent
\textbf{Problem}: Any robot must ensure it avoids collision with other robots and keep in range the neighbour it chooses to follow. Since robots do not know their neighbours views, they cannot predict the moves other robots are about to make. Thus any robot must choose its move so that collisions are avoided and connection with the followed neighbour is preserved for any possible simultaneous move of the other robots.

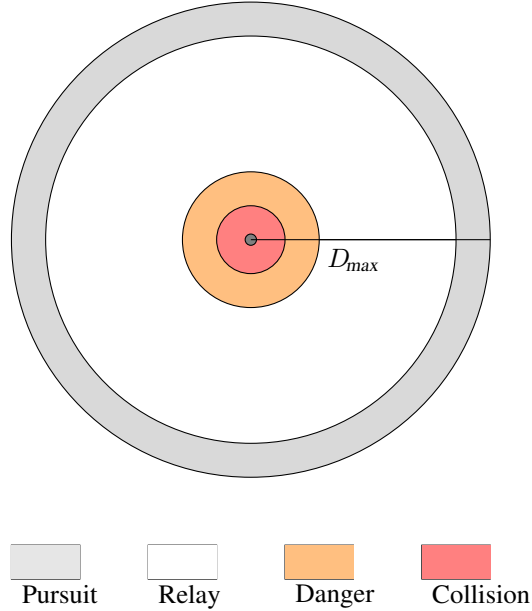
\begin{figure}[!htb]
  \begin{center}
    \begin{tikzpicture}[scale=.6]
      
      \begin{scope}
        \fill (0,0) circle (5.25) [gray!30];
        \fill (0,0) circle (4.5) [white];
        \fill (0,0) circle (4.5) [white];
        \fill (0,0) circle (1.5) [orange!50];
        \fill (0,0) circle (.75) [red!50];
      \end{scope}

        \begin{scope}[xshift=3.75cm,scale=1.5,yshift=1cm]
        \draw (-6,-6) rectangle (-5,-5.5) node [below left=.4cm and
        -.36cm] {Pursuit};
        \fill (-6,-6) rectangle (-5,-5.5) [gray!20];
        \draw (-4,-6) rectangle (-3,-5.5) node [below left=.4cm and
        -.18cm] {Relay};
        \fill (-4,-6) rectangle (-3,-5.5) [white];
        \draw (-2,-6) rectangle (-1,-5.5) node [below left=.4cm and
        -.398cm] {Danger};
        \fill (-2,-6) rectangle (-1,-5.5) [orange!50];
        \draw (-0,-6) rectangle (1,-5.5) node [below left=.4cm and
        -.65cm] {Collision};
        \fill (0,-6) rectangle (1,-5.5) [red!50];
        \end{scope}
        
      \begin{scope}[scale=0.1]
        \robot{0}{0}{black!50}
      \end{scope}
      
      \draw (0,0) circle (.75);
      \draw (0,0) circle (1.5);
      \draw (0,0) circle (4.5);
      \draw (0,0) circle (5.25);
      \draw (4.5,0) -- (5.25,0); 
      \draw (0,0) -- (4.5,0) node [midway, below] {\dmax};
      \draw (1.5,0) -- (4.5,0);
    \end{tikzpicture}
    \end{center}
    \caption{The visible surroundings of a robot can be divided into several
      zones that characterize the situations it may encounter: need to
      follow a robot (Pursuit zone, width $D$), high risk of immediate collision
      (Collision zone, radius $D$) or possibly at the next round (Danger
      zone, width $D$), and finally just transmission (Relay).}

\label{fig:zones}
\end{figure}

\noindent
\textbf{Solution}: Since we assume no robot can travel a distance greater than a constant $D$ in one cycle, we take all decisions with a $D$ margin. This entails defining several zones around a robot using this distance $D$, summed up in Figure~\ref{fig:zones}.
\begin{itemize}
\item A neighbour at distance less than $D$ from the range limit, that
  is a neighbour at a distance comprised between $\dmax-D$ and $\dmax$
  might become out of range after its move. If the followed neighbour
  is in this particular range, a move is necessary to ensure
  connection. This defines the Pursuit Zone. The distance of pursuit
  is then defined as $D_p = \dmax - D$.
\item A neighbour at distance less than $D$ might provoke a collision. Since cycles are atomic we do not model how robots move inside a cycle and we can never ensure that robots with crossing trajectories will not collide. Thus the distance $D$ defines a Collision Zone. When two robots are at a distance less than $D$, either they have to agree on moving apart, or one has to withdraw.
\item A neighbour that is not in the Collision Zone, but is still at a distance less than $2D$ might enter the Collision Zone after its move. This range defines the Danger Zone, and letting another robot enter this zone means risking an imminent collision.
\end{itemize}
The range between distances $2D$ and $\dmax-D$ gives no particular constraint: neighbours in this Relay Zone are neither too close nor too far.

\subsubsection{Robot vision radius must be large enough w.r.t. to
  robots speed}~

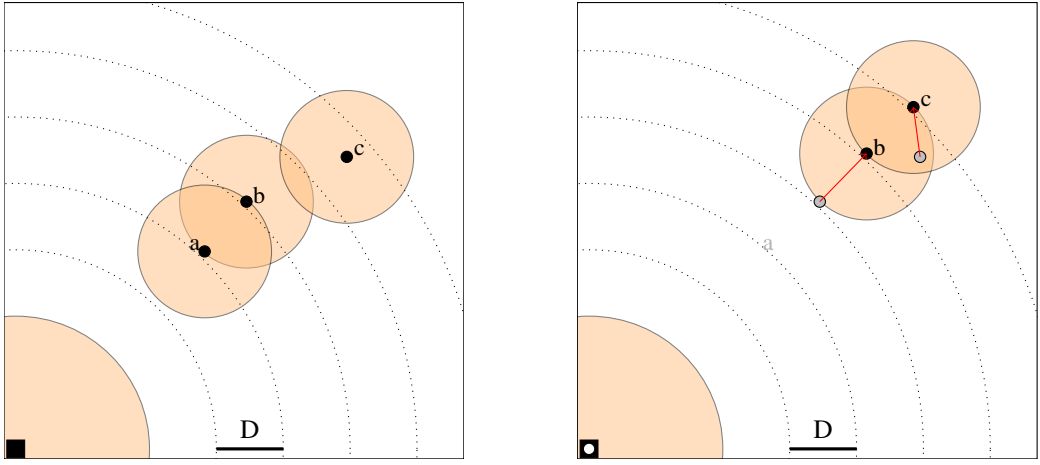
\begin{figure}
\begin{minipage}{.4\linewidth}
  \begin{tikzpicture}[scale=.55]
    \begin{scope}[scale=.4]
      \clip (-7.5,-7.5) rectangle (20,20);
      \draw  (-7.5,-7.5) rectangle (20,20);

    \node[quadri][fill=black] (base) at (-6.8,-6.9){};
    \robotn{7}{8}{black}{r1}
    \node at (r1.north east){\small ~~b};
    \robotn{4.5}{5.0}{black}{r2}
    \node at (r2.north west){\small a~~};
    \robotn{13}{10.7}{black}{r3}
    \node at (r3.north east){\small ~~c};
    \begin{pgfonlayer}{bg} 
      \clip (-7.5,-7.5) rectangle (20,20);
      \draw[opacity = 0.5,fill=orange!50] (base) circle(8);
      \draw[opacity = 0.5,fill=orange!50] (r1) circle(4);
      \draw[opacity = 0.5,fill=orange!50] (r2) circle(4);
      \draw[opacity = 0.5,fill=orange!50] (r3) circle(4);
      \draw[dotted] (base) circle (4*3);
      \draw[dotted] (base) circle (4*4);
      \draw[dotted] (base) circle (4*5);
      \draw[dotted] (base) circle (4*6);
      \draw[dotted] (base) circle (4*7);
      
      \draw[very thick] (-6.8+12,-6.9) -- (-6.8+16,-6.9) node
      [midway, above] {D};
    \end{pgfonlayer}
    \end{scope}
\end{tikzpicture}
\end{minipage}
\hspace{1cm}
\begin{minipage}{.4\linewidth}
  \begin{tikzpicture}[scale=.55]
    \begin{scope}[scale=.4]
      \clip (-7.5,-7.5) rectangle (20,20);
      \draw (-7.5,-7.5) rectangle (20,20);
    \node[quadri][fill=black] (base) at (-6.8,-6.9){};
    \robotn{7}{8}{gray!50}{r1_old}
    \robotn{9.8}{10.9}{black}{r1}
    \draw [->, color=red] (r1_old.center) -- (r1.center);
    \node at (r1.north east){\small ~~b};
    \node at (r2.north west){\textcolor{black!30}{\small a~}};
    \robotn{12.6}{13.7}{black}{r3}
    \robotn{13}{10.7}{gray!50}{r3_old}
    \draw [->, color=red] (r3_old.center) -- (r3.center);
    \node at (r3.north east){\small ~~c};
    \robotn{-6.8}{-6.9}{white}{r0}
    \begin{pgfonlayer}{bg} 
      \clip (-7.5,-7.5) rectangle (20,20);
      \draw[opacity = 0.5,fill=orange!50] (base) circle(8);
      \draw[opacity = 0.5,fill=orange!50] (r1) circle(4);
      \draw[opacity = 0.5,fill=orange!50] (r3) circle(4);
      \draw[dotted] (base) circle (4*3);
      \draw[dotted] (base) circle (4*4);
      \draw[dotted] (base) circle (4*5);
      \draw[dotted] (base) circle (4*6);      
      \draw[dotted] (base) circle (4*7);
      \draw[very thick] (-6.8+12,-6.9) -- (-6.8+16,-6.9) node
      [midway, above] {D};
    \end{pgfonlayer}
\end{scope}
\end{tikzpicture}
\end{minipage}
\caption{
  To avoid elimination at launch, the base waits for a robot a to be
  at more than \dmax$ - 3 D$ before launching another robot. On
  the left, as the base decides to launch a new robot, robot a is too
  close to robot b and will withdraw.  The next round is on the right:
  b is now too close to c and may withdraw, leaving only c, which
  moves further away. As the new robot is launched, robot c \emph{should however
    be still visible from the base}. The new robot \protect\begin{tikzpicture}
       \protect\robot{0}{0}{white}\protect\end{tikzpicture} was launched too late. 
}

  \label{fig:visbound}
 \end{figure}

\noindent
\textbf{Problem}: A robot may not see far enough to connect without
risk. An exemple of this situation is given on
Figure~\ref{fig:visbound} where too long a wait to avoid collision
risks at launch time leads to a connection break.

\noindent
\textbf{Solution}: We put a defensive safe limit for launching
distance at $\dmax - 4D$, small enough to avoid losing robots in the
example. It must also be large enough to allow for safe launch, at
least $3D$. A bound follows on \dmax that must be strictly greater that
$7 D$.

\subsection{A family of solutions}\label{sec:family}

At this point, we have gathered a few conditions that \emph{appear to solve}
the problems we foresee, and should help in maintaining the invariant.
As we formalized them in the \pactole framework, we may now
\emph{prove formally} that indeed any protocol fulfilling these
conditions is a solution to our connection problem.
We define that way a family of protocols satisfying our needs.

Let us first recap the description of this family of solutions before moving to the proof of its correctness.

\subsubsection*{Description of the family of solutions}
We describe the family of solutions to the problem by exhibiting a
generic protocol \lstinline{rbg_fnc} parameterized by three auxiliary functions.
\begin{coqcode}
Context {choose_target : obs_ILA -> (R2*ILA)}.
Context {choose_new_pos: obs_ILA -> location -> location}.
Context {move_to: obs_ILA -> location -> bool }.

Definition rbg_fnc (s:obs_ILA) : R2*light :=
 (* Chose target and new position accordingly *)
 let target := choose_target s in
 let new_pos := choose_new_pos s (fst target) in
 match move_to s new_pos with(*@\hfill@*)(* Is this dangerous? *)
 | true => (new_pos,false)(*@\hfill@*)(* Safe: move + light off. *)
 | false => ((0,0), true)(*@\hfill@*)(* Danger: stay + light on.  *)
 end.
\end{coqcode}

The role of \lstinline!choose_target! is to select among the visible
other robots the one that should be followed, i.e. the one
we should maintain connection with on the next round. The role of
\lstinline!choose_new_pos! is to decide, given the target computed by
\lstinline!choose_target!, where the robot should move to ensure
connection to it.
Finally, the role of \lstinline!move_to! is to decide whether
it is dangerous to move to the selected new position.
If that is the case, then the protocol decides not to move and warns neighbours that it
may withdraw soon by turning on its light (and therefore should preferably not be selected as
a target by another robot on the next round).

Our claim is that any three functions verifying the spe\-ci\-fi\-ca\-tions
described below will make the protocol achieve our goal.
In the following section we give an example of a working instance,
thus ensuring that this family is not empty.

The first hypothesis specifies the behaviour of the \lstinline{choose_target} function.\footnote{In the actual code, each of these five properties is split into its own independent statement for greater flexibility.}
It expresses that its output must:
\begin{itemize}
  \item be in range, that is be within the input (an observation) to \lstinline{choose_target};
  \item be alive;
  \item have a smaller identifier that the observing robot;
  \item preferably have its light \lstinline{off}, that is, if the target has its light on, then all robots within range also do;
  \item preferably select close robots, that is, if the target is in the pursuit zone and has its light on, then all robots within range are also in the pursuit zone.
\end{itemize}
\begin{coqcode}[gobble=2]
  Axiom choose_target_spec : forall obs_id local_config,
    let obs := obs_from_config local_config in
    let target := choose_target obs_id obs in
    (* the target must be in range *)
    target $\in$ obs
    (* the target must be alive *)
    /\ get_alive target = true
    (* the target must have a smaller id *)
    /\ get_ident target < get_ident obs_id
    (* the target must preferably have its light off *)
    /\ (get_light target = true
        -> forall id $\in$ obs, get_light id = true)
    (* the target must preferably be close *)
    /\ (get_light target = true
        -> dist (0,0) (get_loc target) > Dp
        -> forall id $\in$ obs, dist (0,0) (get_loc elt) > Dp).
\end{coqcode}

The second property specifies \lstinline{choose_new_pos}: the new
location is reachable with the speed of the observing robot (it is at
most $D$ away from its current location) and not too far from its
target (at most $D_p$ away from it, thus out of the pursuit zone).
\begin{coqcode}[gobble=2]
  Axiom choose_new_pos_spec : forall obs target,
    let new := choose_new_pos obs target in
    dist new target <= Dp /\ dist new (0,0) <= D.
\end{coqcode}

Finally, the specification of \lstinline{move_to} is split in two depending on whether movement is possible or not.
If it is possible, the chosen location is at least $2D$ away from all other robots. 
\begin{coqcode}[gobble=2]
  Axiom move_to_true_spec : forall obs choice,
    move_to obs choice = true
     -> forall id, id $\in$ obs ->  dist choice get_loc id > 2*D.
\end{coqcode}

If movement is not possible, there is a robot with smaller identifier too close to the chosen location, that is, inside its danger zone (within a disc of radius $2D$).
\begin{coqcode}[gobble=2]
  Axiom move_to_false_spec : forall state local_config new_loc,
    let obs := obs_from_config local_config state in
    move_to obs new_loc = false
     -> exists other, other $\in$ obs
                /\ get_ident other < get_ident state
                /\ dist (get_loc other) new_loc <= 2*D.
\end{coqcode}

\subsubsection*{Correctness proof of the family of solutions}
The proof amounts to ensuring that the property \lstinline{NoCollAndPath} given in Section~\ref{sec:spec} is an invariant of the execution.
This property combines \lstinline{no_collision_conf} (the absence of collision) and \lstinline{path_conf} (which entails the existence of a path from the base to the companion).
We only sketch the proofs to convey their insight, for further detail
we direct the interested reader to the \coq files.

The absence of collision can be directly proven to be an invariant.
Assume that two robots that do not start at the same location collide after their move. Let \lstinline{id} and \lstinline{id}$'$ be the identifiers of the robots, with \lstinline{id} $<$ \lstinline{id}$'$. Since the robots are going to collide and none of them can travel more than $D$, they start at most $2D$ apart.
Since \lstinline{id}$'$ observes a robot with lower identifier \lstinline{id} in its danger zone, it does not move.
Thus, to provoke a collision, \lstinline{id} and \lstinline{id}$'$ should start at most $D$ apart. In this case, \lstinline{id} is actually in the collision zone of \lstinline{id}$'$, so \lstinline{id}$'$ withdraws and there is no collision.

Proving that the existence of a path is invariant is more subtle and we need to introduce three other 
properties
that give more insight into the behaviour of the protocol:
\begin{itemize}
  \item \lstinline{executed_means_light_on} expressing that if a robot withdraws at the next round, then it has its light \lstinline{on} (it is aware that it is in a potentially dangerous situation);
  \item \lstinline{executioner_means_light_off} expressing that if a robot withdraws at the next round, then the robot causing this removal has its light \lstinline{off} (ensuring that it cannot disappear too);
  \item \lstinline{exists_at_less_than_Dp} expressing that if all robots of lower identifier in range of an alive robot $r$ have their light \lstinline{on}, then one of them is not in the pursuit zone of $r$ (at most $D_p$ away from $r$).
\end{itemize}

These properties are proven by case analysis over 
two consecutive
rounds, let us call them \lstinline{cf}, \lstinline{cf}$'$. 

For the property \lstinline{executed_means_light_on}, remark that if a robot $r$ alive in \lstinline{cf} withdraws in \lstinline{cf}$'$, it is either because, in \lstinline{cf}$'$, there is no robot in range or there is one too close (at most $D$ away).
By contradiction, let us assume that $r$ withdraws in
\lstinline{cf}$'$ while not deciding to turn its light \lstinline{on}
in \lstinline{cf}. Since the light is \lstinline{on} when \lstinline{move_to} returns \lstinline{false}, we know this function returns \lstinline{true} in \lstinline{cf}, and $r$ thus performs the move chosen by \lstinline{choose_new_pos} between \lstinline{cf} and \lstinline{cf}$'$. Thus $r$ cannot lose contact with its target: by the specification of \lstinline{choose_new_pos} $r$ moved at a distance no greater than $D_p$ to the \lstinline{cf}-location of its target, that is at a distance no greater than $D_P+D=\dmax$ to the \lstinline{cf}$'$-location of the target. Moreover, $r$ cannot withdraw due to another robot at a distance less than $D$ in \lstinline{cf}$'$: by the specification of \lstinline{choose_new_pos} it moved at a distance more than $2D$ apart from the \lstinline{cf}-location of any other robot of lower identifier, that is more than $2D-D=D$ apart from the \lstinline{cf}$'$-location of any other robot of lower identifier.

For the property \lstinline{executioner_means_light_off},
note that by the definition of the abstract protocol, a moving robot
always has its light \lstinline{off} (by a simple case analysis on
\lstinline{move_to}).

By the property
\lstinline{executed_means_light_on} a robot $r$ alive in
\lstinline{cf} that withdraws in \lstinline{cf}$'$ has its light
\lstinline{on} and did not move between \lstinline{cf} and
\lstinline{cf}$'$. If $r$ stays alive in \lstinline{cf} and withdraws
in \lstinline{cf}$'$, then some other robot $r'$ is at a distance from $r$ that is greater than $D$ in \lstinline{cf} and at most $D$ in \lstinline{cf'}. Since $r$ does not move between the configurations, $r'$ necessary does, and thus has its light \lstinline{off}.

We show now that \lstinline{exists_at_less_than_Dp} holds for \lstinline{cf}$'$. Let us consider an alive robot $r$ in \lstinline{cf}$'$ such that all robots in range have their light \lstinline{on}.
We want to prove that at least one of them is at most $D_p$ away.
Note that as $r$ is alive in \lstinline{cf}$'$, there was a robot in range in \lstinline{cf} and $r$ had a target $r'$ in \lstinline{cf}.
Since $r'$ has its light \lstinline{on} in \lstinline{cf}$'$, it did not move between \lstinline{cf} and \lstinline{cf}$'$.
If $r'$ was out of the pursuit zone of $r$ in \lstinline{cf} (that is, at most $D_p$ away from $r$), we can conclude because $r$ either did not move or moved to a position not farther that $D_p$ away from $r'$.
If $r'$ was inside the pursuit zone if $r$ in \lstinline{cf}, by the specification of \lstinline{choose_target}, so were all robots in range of $r$.
In particular, $r$ could move towards $r'$ as no robot was inside its
danger zone 
and since $r'$ did not move, $r$ and $r'$ are at most $D_p$ away in \lstinline{cf}$'$.

Finally, let us turn to the proof that the property \lstinline{path_conf} is an invariant.
Let $r$ be a relay robot.
We prove that it has a visible, active, launched neighbour with a smaller id.
We consider several cases depending on the value of
\lstinline{move_to} and whether the target of $r$ withdraws.

If \lstinline{move_to} is \lstinline{true} and $r'$ has its light
\lstinline{off}, then by property \lstinline{executioner_means_light_off} $r'$ cannot withdraw, and cannot get out of range of $r$ since $r$ moves closer to $r'$.

If \lstinline{move_to} is \lstinline{true} and $r'$ has its light \lstinline{on}, the invariant \lstinline{exists_at_less_than_Dp} and the specification of \lstinline{choose_target} entail that all robots in range of $r$ have their light \lstinline{on} and that $r'$ is out of the pursuit zone.
Hence, $r'$ cannot withdraw since all robot close enough to eliminate it have their light \lstinline{on} thus cannot cause it to withdraw (by \lstinline{executioner_means_light_off}).

If \lstinline{move_to} is \lstinline{false}, then by \lstinline{move_to_false_spec}, there is a robot with smaller id in range of $r$.

\subsubsection*{Proof effort}

The proof effort for this work is decomposed in setup, specifications
and actual proofs.

It consists of \textbf{520} lines of \emph{setup} to instantiate the context (definition of robots state, observation, etc.), \textbf{540} lines of \coq of \emph{specifications} to describe the problem.

The \emph{proof} part is much more verbose, the table below show how many lines of
proof tactics were used for the main invariants.

\begin{center}
\begin{tabular}{lr}
  \multicolumn{2}{c}{Stability of main invariants} \\ \hline
  {\tt \small no\_collision\_conf}             & 1000 \\
  {\tt \small executed\_means\_light\_on}      & 2300 \\
  {\tt \small executioner\_means\_light\_off}  &  300 \\
  {\tt \small exists\_at\_less\_than\_Dp}      & 1300 \\
  {\tt \small path\_conf}                      &   50 \\
  Auxiliary results                            & $\approx$ 5400 \\ \hline
  Total for proofs                             & \textbf{10500} \\ \hline
\end{tabular}
\end{center}

\subsection{Extracting a Sample Solution}
\label{sec:extr-sample-solut}
In the previous sections we designed a family of solutions defined by a template robogram function. The template robogram is parameterized by three auxiliary functions, each one being restricted by some axioms. Hence we can obtain a concrete solution by providing for each of the auxiliary functions a concrete definition that is consistent with the corresponding axioms.

It is again possible to develop such a concrete solution in our formal
framework. Indeed, the \pactole library already provides some
instances of concrete algorithms proven correct for other
problems~\cite{courtieu16disc,balabonski19tocs}. However, if the axioms are simple enough
 one can also consider that most of the complex and error-prone
 reasoning on the model has been taken care of, and that a traditional
 pen-and-paper check provides a satisfying level of
 certainty. 

 In this section we exhibit a concrete solution, for which we check the axioms by hand. Please note that the selected solution has no particular property, and is not supposed to be more efficient, or better in any sense than the other members of our family of solutions. The concrete solution is chosen as one of the most straightforward validations of the axioms. In other words, the solution we present is a naive solution (or at least, it is naive \emph{once the axioms have been designed}). Such a solution provides a sanity check of our axioms: it witnesses that our family of solutions is not empty.

\subsubsection*{Concretisation of \lstinline{choose_target}}
 
We take the point of view of a robot with identifier $n$, in a configuration satisfying the invariants of the algorithm. Define $V$ the subset of the observed robots that are
\begin{itemize}
\item alive
\item at a distance at most \dmax
\item with an identifier strictly smaller than $n$
\end{itemize}
If there is at least one robot in $V$ whose light is \lstinline{off}, then select any robot in the subset $V_{\texttt{off}}$ of $V$ containing the robots with lights \lstinline{off}. For instance, select the robot in $V_{\texttt{off}}$ with minimal identifier. If however all robots in $V$ have their lights on, then select any robot in the subset $V_{Dp}$ of $V$ containing the robots at a distance at most $D_p$. For instance, select the robot in $V_{Dp}$ with minimal identifier. Otherwise select any robot in $V$, for instance the robot in $V$ with minimal identifier.

We have to check that this function is defined and satisfies all the axioms. First note that, following the invariant \lstinline{path_conf}, there is at least one robot that is alive, at a distance at most \dmax of $n$ and with an identifier strictly smaller than $n$. Hence the set $V$ is not empty, and our function is defined.
Second, the result of this function is a robot of $V$. As such it is visible, alive and and with an identifier strictly smaller than $n$ (three first parts of axiom \lstinline{choose_target_spec} are satisfied).
Third, the result has its light \lstinline{off} as soon as there is at least one visible robot with its light \lstinline{off}: fourth part of \lstinline{choose_target_spec} is also satisfied.
Finally, when there are only robots with their lights on, the result is a robot at a distance at most $D_p$ if there is one: the fifth and final part of axiom \lstinline{choose_target_spec} is again satisfied. Hence our naive \lstinline{choose_target} function satisfies all the relevant axioms.

\subsubsection*{Concretisation of \lstinline{choose_new_pos}}

We take the point of view of a robot with an already chosen target at distance at most \dmax. One can select the potential move of the robot as follows:
\begin{itemize}
\item if the target robot is at a distance greater than $D_p$, then choose a move of length $D$ toward the target;
\item otherwise, choose a null move.
\end{itemize}
This tentative move aims at a position at distance at most $D$ from the starting position, and at most $D_p$ from the target. It thus complies with the axiom \lstinline{choose_new_pos_spec}.

\subsubsection*{Concretisation of \lstinline{move_to}}

We take the point of view of a robot with a potential move already selected, with some observed configuration. Validation of the potential move can be performed as follows:
\begin{itemize}
\item if there is any other robot at most $2D$ away from the potential
  destination, 
  then invalidate the potential move by returning \lstinline{false};
\item otherwise, validate the potential move by returning \lstinline{true}.
\end{itemize}
This function returns \lstinline{true} only if there is no robot at a
distance to the potential destination less or equal to $2D$: the axiom
\lstinline{move_to_Some_zone} is satisfied. Conversely, the function
returns \lstinline{false} only if there is an observable other robot
at a distance to the potential destination less or equal to $2D$: the axiom \lstinline{move_to_None} is also satisfied.

Finally, the three parameter functions \lstinline{choose_target}, \lstinline{choose_new_pos}, and \lstinline{move_to} all satisfy their respective axioms: they define a proper member of our family of solutions, which is not empty.

\section{Concluding remarks}\label{sec:concl}

In this paper, we demonstrated by example how formal methods, and the
\pactole framework in particular, can help mobile robotic swarm
protocol designers to formally specify, design, and prove their
algorithms are correct, balancing expressivity to tackle practically relevant problems, and formality to preserve the mathematical soundness of software developments.

Of course, proving correct algorithms for new problems is only the first step. A natural second step is to ensure the \emph{implementations} of the algorithms maintain the relevant invariants when actually deployed on real devices. We leave this path for future research.


\begin{thebibliography}{10}

\bibitem{altisen16forte}
K.~Altisen, P.~Corbineau, and S.~Devismes.
\newblock A framework for certified self-stabilization.
\newblock In E.~Albert and I.~Lanese, editors, {\em Formal Techniques for
  Distributed Objects, Components, and Systems - 36th {IFIP} {WG} 6.1
  International Conference, {FORTE} 2016, Held as Part of the 11th
  International Federated Conference on Distributed Computing Techniques,
  DisCoTec 2016, Heraklion, Crete, Greece, June 6-9, 2016, Proceedings}, volume
  9688 of {\em Lecture Notes in Computer Science}, pages 36--51.
  Springer-Verlag, 2016.

\bibitem{auger13sss}
C.~Auger, Z.~Bouzid, P.~Courtieu, S.~Tixeuil, and X.~Urbain.
\newblock {Certified Impossibility Results for Byzantine-Tolerant Mobile
  Robots}.
\newblock In T.~Higashino, Y.~Katayama, T.~Masuzawa, M.~Potop-Butucaru, and
  M.~Yamashita, editors, {\em Stabilization, Safety, and Security of
  Distributed Systems - 15th International Symposium (SSS 2013)}, volume 8255
  of {\em Lecture Notes in Computer Science}, pages 178--186, Osaka, Japan,
  Nov. 2013. Springer-Verlag.

\bibitem{balabonski2019netys}
T.~Balabonski, P.~Courtieu, R.~Pelle, L.~Rieg, S.~Tixeuil, and X.~Urbain.
\newblock Continuous vs. discrete asynchronous moves: {A} certified approach
  for mobile robots.
\newblock In M.~F. Atig and A.~A. Schwarzmann, editors, {\em Networked Systems
  - 7th International Conference, {NETYS} 2019, Marrakech, Morocco, June 19-21,
  2019, Revised Selected Papers}, volume 11704 of {\em Lecture Notes in
  Computer Science}, pages 93--109. Springer-Verlag, 2019.

\bibitem{balabonski19tocs}
T.~Balabonski, A.~Delga, L.~Rieg, S.~Tixeuil, and X.~Urbain.
\newblock Synchronous gathering without multiplicity detection: A certified
  algorithm.
\newblock {\em Theory of Computing Systems}, 2019.
\newblock {\scriptsize{\url{https://doi.org/10.1007/s00224-017-9828-z}}}.

\bibitem{balabonski18icdcn}
T.~Balabonski, R.~Pelle, L.~Rieg, and S.~Tixeuil.
\newblock A foundational framework for certified impossibility results with
  mobile robots on graphs.
\newblock In P.~Bellavista and V.~K. Garg, editors, {\em Proceedings of the
  19th International Conference on Distributed Computing and Networking,
  {ICDCN} 2018, Varanasi, India, January 4-7, 2018}, pages 5:1--5:10. {ACM},
  2018.

\bibitem{berard16dc}
B.~B\'{e}rard, P.~Lafourcade, L.~Millet, M.~Potop-Butucaru, Y.~Thierry-Mieg,
  and S.~Tixeuil.
\newblock Formal verification of mobile robot protocols.
\newblock {\em Distributed Computing}, 29(6):459--487, 2016.

\bibitem{bezem97fac}
M.~Bezem, R.~Bol, and J.~F. Groote.
\newblock {Formalizing Process Algebraic Verifications in the Calculus of
  Constructions}.
\newblock {\em Formal Aspects of Computing}, 9:1--48, 1997.

\bibitem{bonnet14wssr}
F.~Bonnet, X.~D{\'{e}}fago, F.~Petit, M.~Potop{-}Butucaru, and S.~Tixeuil.
\newblock Discovering and assessing fine-grained metrics in robot networks
  protocols.
\newblock In {\em 33rd {IEEE} International Symposium on Reliable Distributed
  Systems Workshops, {SRDS} Workshops 2014, Nara, Japan, October 6-9, 2014},
  pages 50--59. {IEEE}, 2014.

\bibitem{castenow20spaa}
J.~Castenow, P.~Kling, T.~Knollmann, and F.~M. auf~der Heide.
\newblock A discrete and continuous study of the max-chain-formation problem:
  Slow down to speed up.
\newblock In C.~Scheideler and M.~Spear, editors, {\em {SPAA} '20: 32nd {ACM}
  Symposium on Parallelism in Algorithms and Architectures, Virtual Event, USA,
  July 15-17, 2020}, pages 515--517. {ACM}, 2020.

\bibitem{coquand90colog}
T.~Coquand and C.~Paulin-Mohring.
\newblock {Inductively Defined Types}.
\newblock In P.~Martin-L{\"o}f and G.~Mints, editors, {\em {International
  Conference on Computer Logic ({C}olog'88)}}, volume 417 of {\em Lecture Notes
  in Computer Science}, pages 50--66. Springer-Verlag, 1990.

\bibitem{courtieu15ipl}
P.~Courtieu, L.~Rieg, S.~Tixeuil, and X.~Urbain.
\newblock {Impossibility of Gathering, a Certification}.
\newblock {\em Information Processing Letters}, 115:447--452, 2015.

\bibitem{courtieu16disc}
P.~Courtieu, L.~Rieg, S.~Tixeuil, and X.~Urbain.
\newblock Certified universal gathering algorithm in {$\mathbb{R}^2$} for
  oblivious mobile robots.
\newblock In C.~Gavoille and D.~Ilcinkas, editors, {\em Distributed Computing -
  30th International Symposium, (DISC 2016)}, volume 9888 of {\em Lecture Notes
  in Computer Science}, Paris, France, Sept. 2016. Springer-Verlag.

\bibitem{cousineau12fm}
D.~Cousineau, D.~Doligez, L.~Lamport, S.~Merz, D.~Ricketts, and H.~Vanzetto.
\newblock {TLA} + {P}roofs.
\newblock In D.~Giannakopoulou and D.~M{\'e}ry, editors, {\em FM}, volume 7436
  of {\em Lecture Notes in Computer Science}, pages 147--154, Paris, France,
  Aug. 2012. Springer-Verlag.

\bibitem{defago20srds}
X.~D{\'{e}}fago, A.~Heriban, S.~Tixeuil, and K.~Wada.
\newblock Using model checking to formally verify rendezvous algorithms for
  robots with lights in euclidean space.
\newblock In {\em International Symposium on Reliable Distributed Systems,
  {SRDS} 2020, Shanghai, China, September 21-24, 2020}, pages 113--122. {IEEE},
  2020.

\bibitem{deng09tase}
Y.~Deng and J.-F. Monin.
\newblock {Verifying Self-stabilizing Population Protocols with Coq}.
\newblock In W.-N. Chin and S.~Qin, editors, {\em Third IEEE International
  Symposium on Theoretical Aspects of Software Engineering (TASE 2009)}, pages
  201--208, Tianjin, China, July 2009. IEEE Computer Society.

\bibitem{devismes12sss}
S.~Devismes, A.~Lamani, F.~Petit, P.~Raymond, and S.~Tixeuil.
\newblock {Optimal Grid Exploration by Asynchronous Oblivious Robots}.
\newblock In A.~W. Richa and C.~Scheideler, editors, {\em Stabilization,
  Safety, and Security of Distributed Systems - 14th International Symposium
  (SSS 2012)}, volume 7596 of {\em Lecture Notes in Computer Science}, pages
  64--76, Toronto, Canada, Oct. 2012. Springer-Verlag.

\bibitem{doan16sofl}
H.~T.~T. Doan, F.~Bonnet, and K.~Ogata.
\newblock Model checking of a mobile robots perpetual exploration algorithm.
\newblock In S.~Liu, Z.~Duan, C.~Tian, and F.~Nagoya, editors, {\em Structured
  Object-Oriented Formal Language and Method - 6th International Workshop,
  {SOFL+MSVL} 2016, Tokyo, Japan, November 15, 2016, Revised Selected Papers},
  volume 10189 of {\em Lecture Notes in Computer Science}, pages 201--219,
  2016.

\bibitem{doan17opodis}
H.~T.~T. Doan, F.~Bonnet, and K.~Ogata.
\newblock Model checking of robot gathering.
\newblock In J.~Aspnes, A.~Bessani, P.~Felber, and J.~Leit{\~{a}}o, editors,
  {\em 21st International Conference on Principles of Distributed Systems,
  {OPODIS} 2017, Lisbon, Portugal, December 18-20, 2017}, volume~95 of {\em
  LIPIcs}, pages 12:1--12:16. Schloss Dagstuhl - Leibniz-Zentrum f{\"{u}}r
  Informatik, 2017.

\bibitem{flocchini05tcs}
P.~Flocchini, G.~Prencipe, N.~Santoro, and P.~Widmayer.
\newblock Gathering of asynchronous robots with limited visibility.
\newblock {\em Theor. Comput. Sci.}, 337(1-3):147--168, 2005.

\bibitem{fokkink07}
W.~Fokkink.
\newblock {\em {Modelling Distributed Systems}}.
\newblock EATCS Texts in Theoretical Computer Science. Springer-Verlag, 2007.

\bibitem{gaspar14ijpp}
N.~Gaspar, L.~Henrio, and E.~Madelaine.
\newblock Bringing coq into the world of gcm distributed applications.
\newblock pages 643--662, 2014.

\bibitem{kling19bookchapter}
P.~Kling and F.~M. auf~der Heide.
\newblock Continuous protocols for swarm robotics.
\newblock In P.~Flocchini, G.~Prencipe, and N.~Santoro, editors, {\em
  Distributed Computing by Mobile Entities, Current Research in Moving and
  Computing}, volume 11340 of {\em Lecture Notes in Computer Science}, pages
  317--334. Springer, 2019.

\bibitem{kuefner12ifiptcs}
P.~K{\"u}fner, U.~Nestmann, and C.~Rickmann.
\newblock {Formal Verification of Distributed Algorithms - From Pseudo Code to
  Checked Proofs}.
\newblock In J.~C.~M. Baeten, T.~Ball, and F.~S. de~Boer, editors, {\em IFIP
  TCS}, volume 7604 of {\em Lecture Notes in Computer Science}, pages 209--224,
  Amsterdam, The Netherlands, Sept. 2012. Springer-Verlag.

\bibitem{lamport1994ACM}
L.~Lamport.
\newblock The temporal logic of actions.
\newblock {\em ACM Trans. Program. Lang. Syst.}, 16(3):872–923, May 1994.

\bibitem{MPST14c}
L.~Millet, M.~Potop{-}Butucaru, N.~Sznajder, and S.~Tixeuil.
\newblock On the synthesis of mobile robots algorithms: The case of ring
  gathering.
\newblock In P.~Felber and V.~K. Garg, editors, {\em Stabilization, Safety, and
  Security of Distributed Systems - 16th International Symposium, (SSS 2014)},
  volume 8756 of {\em Lecture Notes in Computer Science}, pages 237--251,
  Paderborn, Germany, Sept. 2014. Springer-Verlag.

\bibitem{reynaud16ahn}
L.~Reynaud and I.~G. Lassous.
\newblock Design of a force-based controlled mobility on aerial vehicles for
  pest management.
\newblock {\em Ad-Hoc Networks}, 53:41--52, 2016.

\bibitem{sangnier20fmsd}
A.~Sangnier, N.~Sznajder, M.~Potop{-}Butucaru, and S.~Tixeuil.
\newblock Parameterized verification of algorithms for oblivious robots on a
  ring.
\newblock {\em Formal Methods Syst. Des.}, 56(1):55--89, 2020.

\bibitem{suzuki99siam}
I.~Suzuki and M.~Yamashita.
\newblock {Distributed Anonymous Mobile Robots: Formation of Geometric
  Patterns}.
\newblock {\em SIAM Journal of Computing}, 28(4):1347--1363, 1999.

\end{thebibliography}
\end{document}